\documentclass{aa}
\usepackage{graphicx,natbib,amssymb}
\bibpunct{(}{)}{;}{a}{}{,}

\def\1p5{CBF}

\begin{document}

\sloppypar

   \title{Interstellar extinction and the distribution of stellar populations
in the direction of the ultra-deep {\em Chandra} Galactic field}

   \author{M.~Revnivtsev\inst{1,2,3}, M. van den Berg \inst{4},
R.~Burenin\inst{2}, J.E.Grindlay \inst{4}, D. Karasev\inst{2}, W.Forman \inst{4}}


   \institute{
        Excellence Cluster Universe, Technische Universit\"at M\"unchen, Boltzmannstr.2, 85748 Garching, Germany
\and
              Space Research Institute, Russian Academy of Sciences,
              Profsoyuznaya 84/32, 117997 Moscow, Russia
\and
Visiting Fellow, Harvard-Smithsonian Center for Astrophysics, 60 Garden Street, Cambridge, MA 02138, USA
\and
Harvard-Smithsonian Center for Astrophysics, 60 Garden Street, Cambridge, MA 02138, USA
 }
  \date{}
\authorrunning{Revnivtsev et al.}
\titlerunning{Extinction and distribution of stars in the ultra-deep
{\em Chandra} galactic field}

\abstract{We studied the stellar population in the central 6.6\arcmin
$\times$ 6.6\arcmin\,region of the ultra-deep (1Msec) {\em Chandra}
Galactic field--- the ``Chandra bulge field'' (CBF) approximately 1.5 
degrees away from the Galactic Center ---using the {\em Hubble Space
Telescope} ACS/WFC blue (F435W) and red (F625W) images. We mainly
focus on the behavior of red clump giants -- a distinct stellar
population, which is known to have an essentially constant intrinsic
luminosity and color. By studying the
variation in the position of the red clump giants on a
spatially resolved color-magnitude diagram, we confirm the anomalous
total-to-selective extinction ratio, as reported
in previous work for other Galactic bulge fields.  We show that the
interstellar extinction in this area is $\langle A_{\rm F625W} \rangle
= 4$ on average, but varies significantly between $\sim3-5$ on angular
scales as small as 1 arcminute. Using the distribution of red clump
giants in an extinction-corrected color-magnitude diagram, we
constrain the shape of a stellar-mass distribution model in
the direction of this ultra-deep {\em Chandra} field, which will be
used in a future analysis of the population of X-ray sources. We also
show that the adopted model for the stellar density distribution
predicts an infrared surface brightness in the direction of
the ``Chandra bulge field'' in good agreement (i.e.~within $\sim$15\%) with
the actual measurements derived from the {\em Spitzer}/IRAC observations. }

   \maketitle

%

\section{Introduction}
Our Galaxy hosts discrete X-ray sources of various types. The X-ray
luminosities of Galactic sources range from $L_x\sim10^{39}-10^{40}$
erg s$^{-1}$ for very luminous accreting neutron star and black hole
binaries down to $10^{26}-10^{30}$ erg s$^{-1}$ for coronally active
stars. Statistical properties of these different populations of
sources are poorly known with large allowed ranges or uncertainties, which in some cases might lead
to an underestimation of their contribution to the global X-ray
emission of our Galaxy and other galaxies.

Constraining the luminosity
functions of the X-ray emitting populations is particularly relevant for
understanding the faint, unresolved X-ray emission of the Galaxy that
is distributed quite smoothly along the Galactic plane (the so called
Galactic ridge X-ray emission, GRXE), which was discovered in
early X-ray experiments \cite[e.g.][]{worrall82}. Until recently,
the nature of the GRXE remained unexplained, to a large extent due to
our scarce knowledge of the statistical properties of populations of
Galactic X-ray sources. Only recently it was shown that at least the
majority of this unresolved X-ray emission arises from the
cumulative emission of a large number of faint discrete sources,
predominantly accreting white dwarfs and coronally active stars
\citep{revnivtsev06,revnivtsev09}.

Understanding the origin of the GRXE would have been impossible
without the progress in understanding the statistical properties of
faint Galactic X-ray sources, which was achieved through
X-ray all-sky surveys \citep{sazonov06}. However, due to the limiting
sensitivity of currently available all-sky surveys, these studies
can probe only sources near the Sun, and the number of sources that
are used to construct luminosity functions is small. This leads to
significant uncertainties in the inferred properties and leaves the
question about possible variations of populations throughout the
Galaxy open. The resulting uncertainties have direct consequences
for our studies of distant galaxies. As has been shown, the
GRXE-type emission is a dominant component (after subtracting the contribution from
low mass and high mass X-ray binaries with luminosities $L >10^{35}$ erg sec$^{-1}$) for a significant fraction of
non-starburst galaxies \cite[see e.g.][]{revnivtsev07,revnivtsev08}. Therefore
it is important to understand the level of universality of the X-ray
luminosity per unit solar mass
inferred from studies of nearby objects. Another non-local sample of
faint Galactic X-ray sources is required to address these questions.

The brightest ($L_x\gtrsim10^{32}$ erg s$^{-1}$) end of the luminosity
functions of various classes of faint Galactic sources can be probed
over distances of several kilo-parsecs by surveys of the plane with a
limiting sensitivity $F_x\approx10^{-15}-10^{-14}$ erg s$^{-1}$
cm$^{-2}$ and this is now extensively explored
\cite[e.g.][]{hands04,grindlay05}.  But for fainter sources one needs
to perform much deeper observations (with limiting sensitivity $\sim10^{-16}-10^{-17}$ erg sec$^{-1}$ cm$^{-2}$) and choose a region with
minimal obscuring dust and gas to avoid the effects of interstellar
absorption. These observations with a total exposure time of 1~Msec
were recently performed with the {\em Chandra X-ray Observatory} in a
direction approximately 1.5 degree away from the Galactic center.
This field, located at the galactic coordinates
($l$,$b$)=(0.10$^{\circ}$,$-$1.43$^{\circ}$) and chosen for its low
extinction \cite[$A_V\approx4$; see e.g.][]{drimmel03} and proximity
to the Galactic Center, was first observed with {\em Chandra} in 2005
for 100 ks (``LimitingWindow'' target) to study the nature and radial 
distribution of hard (2--10
keV) X-ray point sources in the inner Galactic bulge \citep{hong09}.
Simultaneously, images with the Advanced Camera for Surveys Wide Field
Camera (ACS/WFC) on the {\em Hubble Space Telescope} ({\em HST}) were
obtained of the central 6.6\arcmin $\times$ 6.6\arcmin\  region (centered at
RA=267.86968 deg, Dec=-29.581033 deg , J2000) of the field
observed by {\em Chandra} to search for optical counterparts
\citep{vandenberg09}. Analogous to other inner-bulge fields included
in this survey (e.g.\, Baade's Window and Stanek's Window, at
$\sim$3.8-degree and $\sim$2.2-degree offsets from the Galactic
center, respectively) this field was called the Limiting Window,
which refers to the rapidly declining opportunities for optical
identification when moving closer to the heavily-obscured Galactic
center region. In 2008, Revnivtsev et al., who referred to this region
as the ``1.5degree field'', indicating its approximate angular distance 
from the
Galactic Center---revisited this field with {\em Chandra} for 900 ks
to derive constraints on the resolved fraction of the GRXE
\citep{revnivtsev09}. Below we refer to this
field as the \1p5 -- ``Chandra Bulge field''.

\begin{figure}[htb]
\includegraphics[width=0.9\columnwidth]{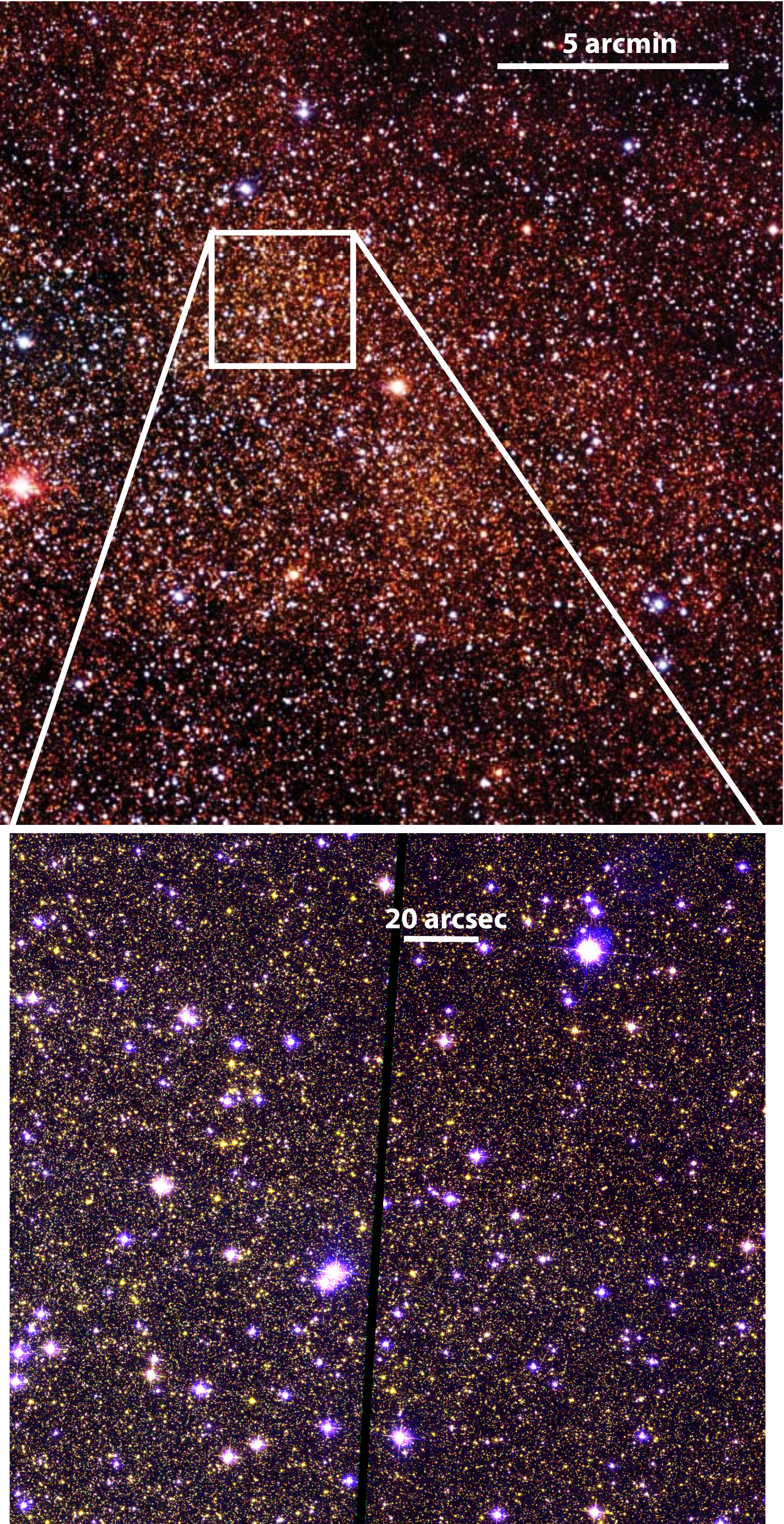}
\caption{Three-color optical images of the field containing the ultra-deep {\em Chandra}
observations of the \1p5. The upper panel shows the
mosaic image obtained by the 1.5m Russian-Turkish telescope (RTT150 telescope) (red shows the {\em i'}
spectral band, green - {\em r'}, blue - {\rm g'}), while the lower
panel shows a smaller area observed with the {\em HST} ACS/WFC (red -
F625W, green - F658N, blue - F435W). From the top image it is evident
that the interstellar extinction is strongly variable across the field
(the regions close to boundaries of the figure are darker than the
central regions). It is seen that the majority of stars are relatively
faint and yellow -- these are the old bulge population of the Galaxy.}
\label{image}
\end{figure}

In comparison with the Solar neighborhood, the analysis of the origin
of the X-ray emission from the \1p5 has its own complications:
\begin{itemize}
\item First of all, we need to know the stellar density enclosed by the volume
studied by {\em Chandra}.
\item We need to understand the importance of the absorption
that results from cold interstellar matter along the
line of sight.
\end{itemize}
To study these issues, we used observations of the \1p5 in the
optical band. In addition to the deep narrow-field images obtained
with {\em HST}, large mosaics ($\sim 25^\prime\times 25^\prime$) of the field were obtained with
the ground-based 1.5m Russian-Turkish telescope RTT150 (see Fig.~\ref{image}) and the
6.5-m Magellan/Baade telescope (see Fig.\ref{comparison_magellan}). In this paper
we concentrate on the properties of the red clump giants
(Sect.~\ref{sec_rcg}) in the
\1p5 as derived from the {\em HST} images
(Sect.~\ref{sec_rcghst}) to infer properties of the interstellar
extinction in this direction (Sect.~\ref{sec_ext}) and constrain
the stellar density distribution along the line of sight
(Sects.~\ref{sec_model} and \ref{sec_distribution}). We also used results of
the measurements with {\em
Spitzer}/IRAC to compare the observed infrared emission towards
the \1p5 with that predicted by the adopted stellar density model
(Sect.~\ref{sec_spitzer}). Our conclusions are summarized in
Sect.~\ref{sec_conclusion}. Other properties of the stellar population
like age and metalicity will be explored in separate works. The
ultimate goal, also to be addressed in a follow-up paper, is to derive
the specific X-ray emission per unit mass and use this to study the
unresolved X-ray emission from other galaxies.

\section{Extinction in the \1p5}

\subsection{Red clump giants} \label{sec_rcg}

\begin{figure}[htb]
\includegraphics[width=\columnwidth]{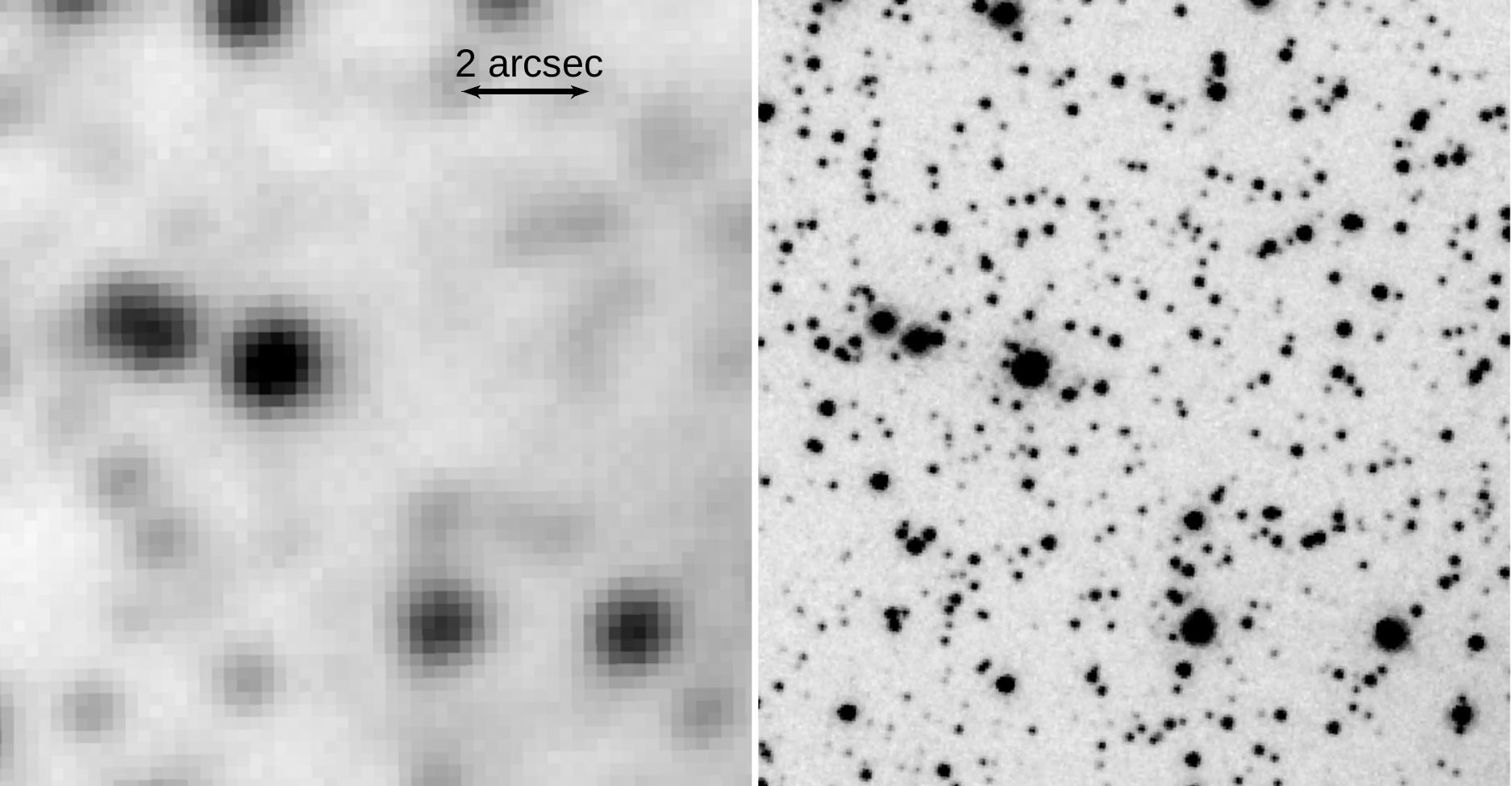}
\caption{This comparison of a small part of the \1p5 as observed by
Magellan/IMACS ($R$) (left) and the {\em HST} ACS/WFC (F625W)
(right) demonstrates the problem of confusion in ground-based images.}
\label{comparison_magellan}
\end{figure}

The red giant branch in the Hertzprung-Russell diagram shows a distinct feature that corresponds to a
special subpopulation known as the red clump giants (RCGs). The
intrinsic colors and absolute magnitudes of these core helium-burning
stars are well-defined with
little dispersion
\cite[e.g.][]{paczynski98} and only a weak dependence on age and
metallicity. Therefore RCGs can play the role of
standard candles and are very useful for various studies of either
stellar populations or properties of the interstellar medium. In
particular, these stars have been used to map the Galactic bar
\citep{stanek94,stanek97}, to measure distances to the Galactic
Center \citep{paczynski98} and to the galaxy M31 \citep{stanek98}, and to
measure the properties \cite[e.g.][]{wozniak96} and spatial variations
\citep{sumi04} of interstellar extinction in the Galactic bulge.
Ground based observations with their limited angular resolution,
encounter severe confusion problems in regions close to the Galactic
Center (GC), which complicates the stellar photometry. Because of this
problem and also due to the very large extinction, the majority of RCG
studies were previously limited to angular distances larger than $\sim$2$^\circ$
from the GC, while our field of interest is located at
$\sim$1.5$^\circ$ from the GC.  As a demonstration of the confusion
problems, we compare images of our field as observed by Magellan/IMACS
with a seeing of 0.8\arcsec\, and by {\em HST} ACS/WFC in
Fig.~\ref{comparison_magellan}. As the figure demonstrates, the ground
based data are strongly
confusion limited. Photometric measurements of stars with magnitudes fainter than $\sim17-18$
are strongly affected by the contribution of fainter unresolved stars.

\subsection{Intrinsic properties of RCGs in the F435W and F625W filters} \label{sec_rcghst}

To understand the luminosity and intrinsic colors of an unabsorbed population of
RCGs in the selected {\em HST}/ACS filters, we used the catalog of
284 {\em Hipparcos} RCGs from
\cite{alves00}. As we are unable to make real observations of all
listed RCGs in these two filters, we calculated their apparent
brightnesses with the Kurucz model atmosphere spectra as included in the
{\tt SYNPHOT/STSDAS/IRAF}
package\footnote{http://www.stsci.edu/resources/software\_hardware/stsdas/
synphot} \citep{bushouse94} as follows.  For all stars in Table~1 of
\cite{alves00} we selected the stellar models that best
match the combination of $B-V$ and $V-I$ colors as presented in
\cite{alves00}. The resulting spectra were then convolved with the
F435W and F625W transmission curves in {\tt SYNPHOT} to calculate the
magnitudes in these filters.  We assumed that all stars in
\cite{alves00} have zero extinction, because they are all very
nearby.  As a check of the overall approach, we calculated the
predicted magnitudes of the stars in the $K$ band and compared them
with their observed $K$ magnitudes as listed in
\cite{alves00}. The difference between the predicted and observed magnitudes has a normal
distribution with zero mean and rms scatter of 0.13 mag, demonstrating
that our approach is reasonable. We might expect that the accuracy of
the predicted brightnesses in the F435W and F625W filters is even
better because their transmissions are closer to those of the $B$, $V$
and $I$ filters---which we used as a reference for the spectral
modeling---than to the $K$-band response.

\begin{figure}
\includegraphics[width=\columnwidth,bb=14 128 581 750,clip]{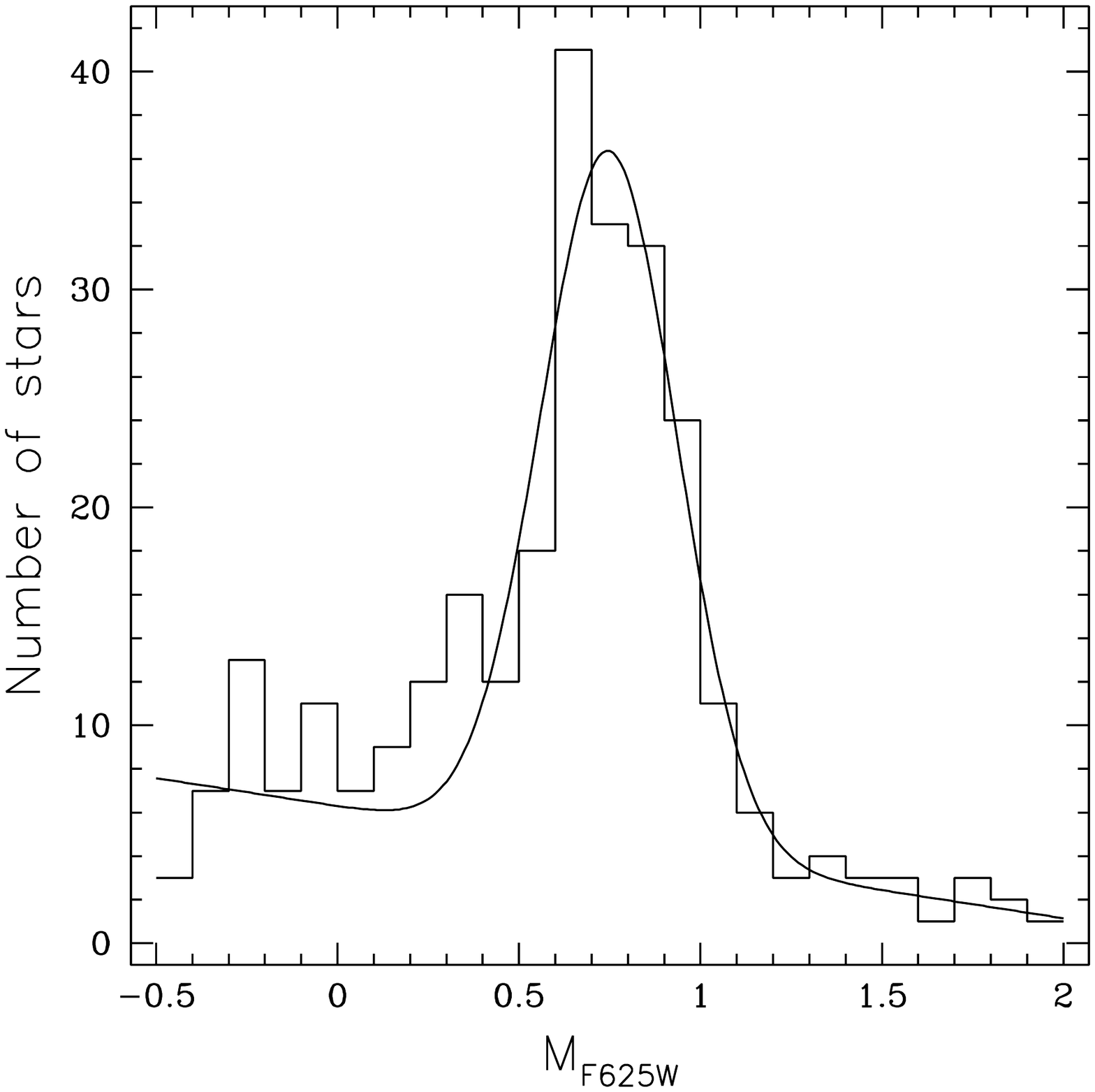}
\caption{Distribution of the absolute brightness of the 284 red clump giants in
Table~1 of \cite{alves00}, converted into the F625W {\em HST}/ACS
filter. The solid curve is the model fit to the distribution. It
consists of a linear function plus a Gaussian with centroid at $M_{\rm
F625W}=0.75$ and $\sigma=0.19$.}
\label{distr_abs}
\end{figure}

The resulting distribution of absolute magnitudes of RCGs in the
F625W filter is presented in Fig.~\ref{distr_abs}. The distribution is
clearly peaked, with centroid at $M_{\rm F625W}=0.75$ and an rms
scatter around this value of $0.19$. The solid line model includes also the 
linear component following the approach of \cite{alves00}, which reflects 
some "background" population of 
red giants in the considered color and brightness intervals.
The color of the peak of the
RCG distribution is $\langle M_{\rm F435W}-M_{\rm F625W}\rangle=0.52$ with an rms
scatter $0.18$. There is no strong indication for the dependence of
the position of the peak on color.

As an additional check, we repeated the above analysis with the {\tt
SYNPHOT} spectra from the Pickles atlas \citep{pickles98} and found the same
distribution for $M_{\rm F625W}$.

\subsection{Ratio of the total-to-selective extinction} \label{sec_ext}

For our subsequent analysis, we used the {\em HST}/ACS images taken
with the F435W and F625W filters, which are somewhat similar to the SDSS
{\it g'}\, and {\it r'}\, filters, to identify the RCGs in the \1p5
and derive their magnitudes. Details of the photometric analysis are
given in \cite{vandenberg09}. Below we will use the STMAG photometric
system.

It has been shown by several authors that the extinction law towards
the Galactic bulge differs from the standard one and
in addition shows variations between different lines of sight
\cite[see e.g.][]{popowski00,udalski03,sumi04}. Therefore  we must
determine the extinction law before we
are able to determine the total extinction in our field. Like \cite{udalski03} and
\cite{sumi04} we used the shift of the RCGs in the
($m_{\rm F625W}$, $m_{\rm F435W}-m_{\rm F625W}$) color-magnitude
diagram due to variable extinction in our field to estimate the ratio
of the total-to-selective extinction. This shift can be clearly seen in
the spatially-resolved ($m_{\rm F625W}$, $m_{\rm F435W}-m_{\rm
F625W}$) color-magnitude diagrams in Fig.~2 of
\cite{vandenberg09}. Then, using the mean intrinsic color of RCGs as
determined above, we calculated the total extinction towards the
\1p5.

\begin{figure}[htb]
\includegraphics[width=\columnwidth,bb=28 172 572 704,clip]{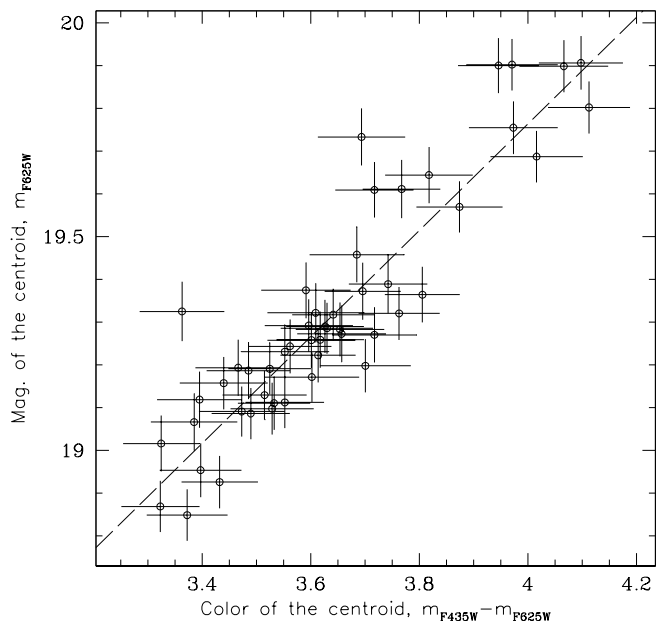}
\caption{Dependence of the mean magnitude of RCGs on their
mean colors for different locations within the
{\em HST}/ACS field. The slope of the linear fit to these
positions (solid line) measures the ratio of the total to selective
extinction $R_{\rm F625W,F435W}=1.25\pm0.09$.}
\label{total_and_selective}
\end{figure}

We divided the area observed by {\em HST}/ACS into rectangular bins of
the size $0.96^\prime\times0.96^\prime$. Of the stars detected in each
angular bin
we filtered out those that are clearly not associated with
the red clump, viz.~those that are fainter in F625W than $m_{\rm
F625W}>17+(m_{\rm F435W}-m_{\rm F625W})$ and have a color $(m_{\rm
F435W}-m_{\rm F625W})$ less than 2.0. To measure the mean magnitude and 
color of RCGs, we then determined the mean color
and magnitude of the stars within an ellipse of the height $\Delta m_{\rm
F625W}=0.5$ mag and width $\Delta (m_{\rm F435W}-m_{\rm F625W})=1$ mag
which is centered on the peak of the number distribution of the
remaining stars in the color-magnitude diagram. The resulting
dependence of the mean magnitude of RCGs on their mean color is
presented in Fig.~\ref{total_and_selective}. The variation of the RCG
mean magnitude with color can be adequately
described by a linear function with a ratio of the total-to-selective
extinction $R_{\rm F625W,F435W} \equiv \Delta
\langle m_{\rm F625W} \rangle /\Delta \langle m_{\rm F435W}-m_{\rm
F625W}\rangle=1.25\pm0.09$. The scatter of the measured centroids of
the RCG positions around this fit is compatible with the statistical
uncertainties in the individual centroids. The total number of RCGs
used in these calculations is on the order of a few hundreds per $0.96^\prime\times0.96^\prime$
bin.  Note that the standard extinction law predicts $R_{\rm
F625W,F435W} \approx 1.8-1.9$
\citep{sirianni05}, thus larger than we measure in the \1p5\ (a similar conclusion 
regarding the smaller ratio of the total-to-selective extinction  was found previously  by e.g. \citealt{popowski00,udalski03,sumi04})

\begin{figure}[htb]
\begin{center}
\vbox{
\includegraphics[width=0.7\columnwidth]{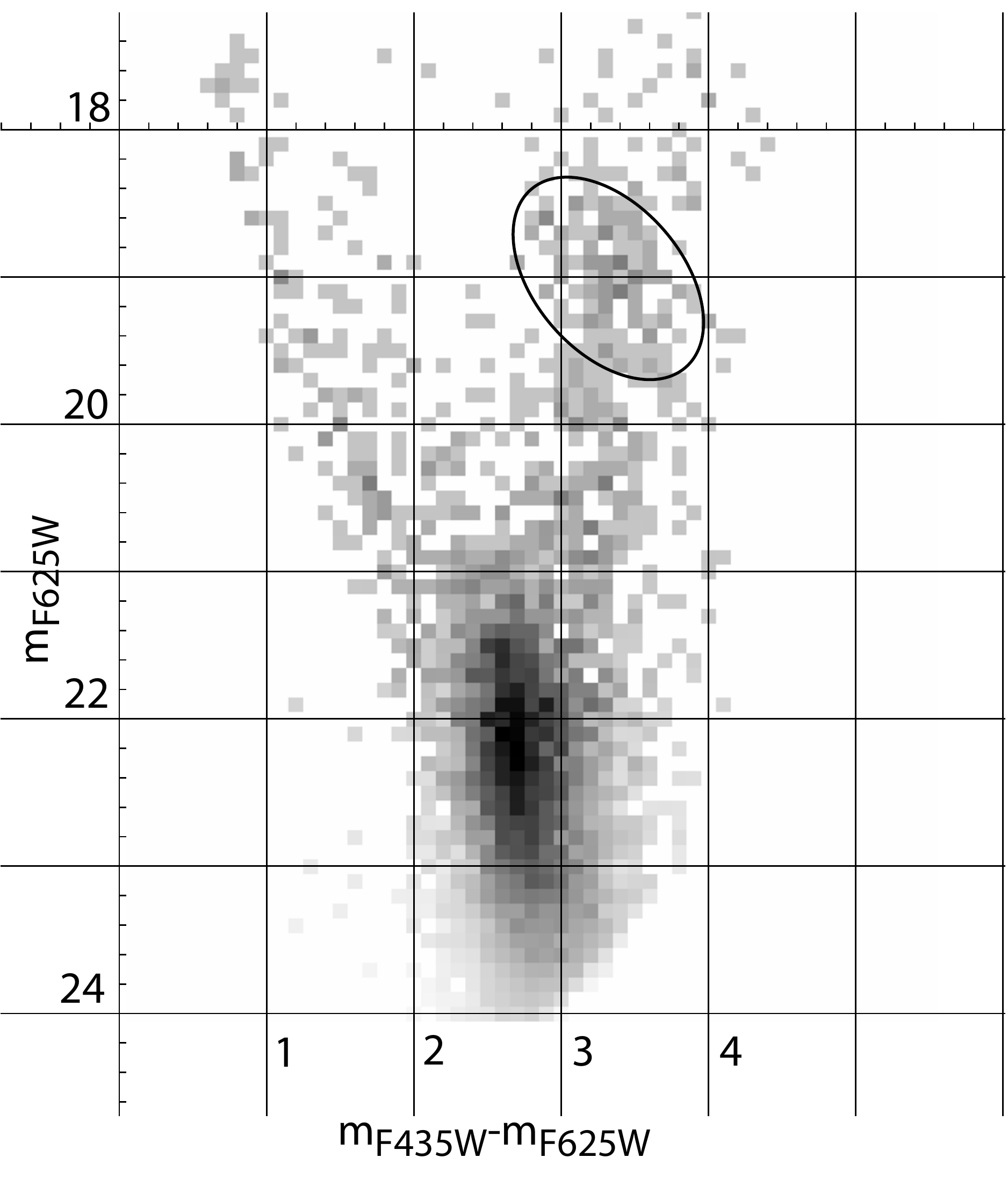}
\includegraphics[width=0.7\columnwidth]{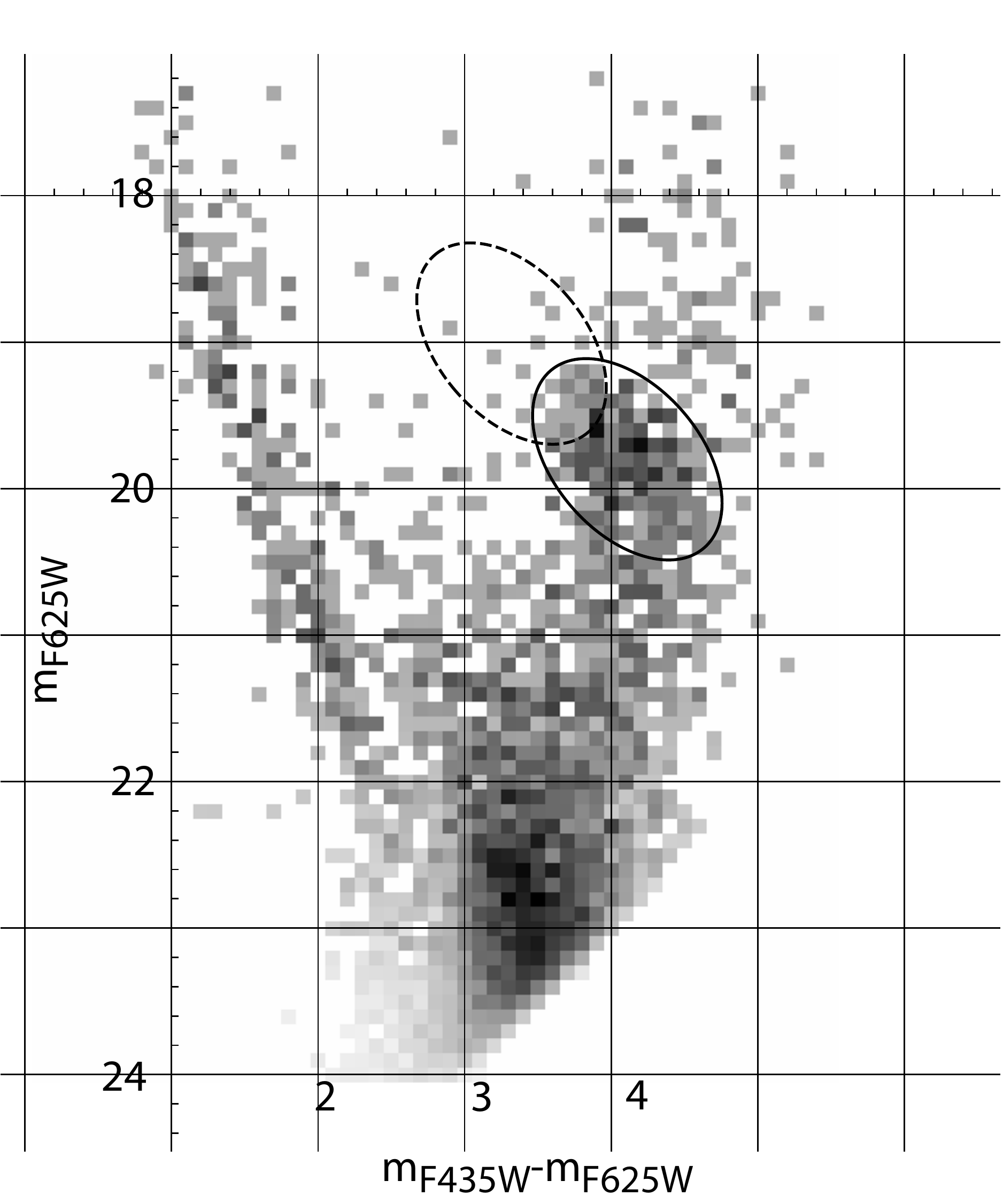}
}
\end{center}
\caption{Color-magnitude diagrams in two areas of the {\em HST}/ACS
field, with the number of stars per image bin color coded.  The
number of faint stars ($m_{\rm F625W}>20.5$) is artificially depressed
via multiplication by a factor $\exp(-(m_{\rm F625W}-20.5)^2/3)$. The
positions of the RCGs are marked with ellipses. In the
lower panel, the ellipse with the dashed outline shows the position of
the RCGs in the color-magnitude diagram of the upper panel. The
ellipses are elongated in the direction of the reddening in the field;
note that this direction is somewhat different from the one predicted
by the standard extinction law, see text.}
\label{two_cmd}
\end{figure}

\begin{figure}[htb]
\includegraphics[width=\columnwidth]{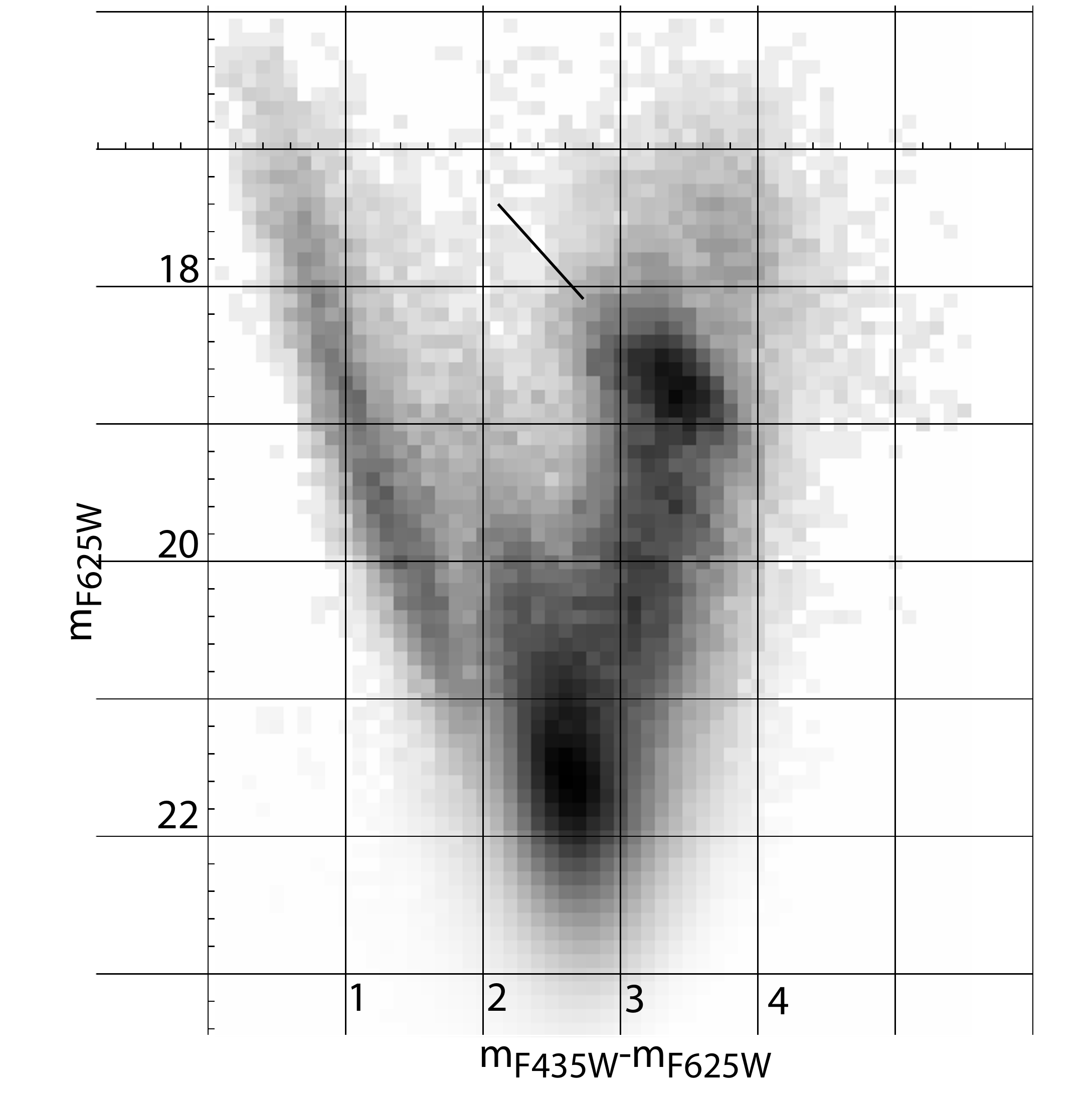}
\caption{Color-magnitude diagram of all photometered stars in the
{\em HST}/ACS field. The magnitudes and colors of stars in each
rectangular $0.96^\prime\times0.96^\prime$ bin are extinction
corrected to the value $A_{\rm F625W}\sim3.5$. The number of faint
stars ($m_{\rm F625W}>19.5$) is artificially depressed via
multiplication by a factor $\exp(-(m_{\rm F625W}-19.5)^2/2)$. The line
shows the direction of the reddening. The number of stars per image
bin is color coded.}
\label{cmd_total}
\end{figure}

The position of the red clump in the color-magnitude diagrams of the two
different spatial bins is shown in Fig.~\ref{two_cmd}. A color-magnitude diagram of
all photometered stars in the {\em HST}/ACS images with the extinction
reduced to the value of $A_{\rm F625W}\sim3.5$ is shown in Fig.~\ref{cmd_total}.

Now we computed the apparent magnitude of RCGs corrected for
extinction ($m_{\rm F625W,0}$) in each spatial bin with the
assumption that their intrinsic mean color is $\langle m_{\rm
F435W}-m_{\rm F625W}\rangle_0=0.52$:
$$
\langle m_{\rm F625W} \rangle = m_{\rm F625W,0} + R_{\rm F625W,F435W} \times
$$
$$
(\langle m_{\rm F435W} - m_{\rm F625W} \rangle - \langle m_{\rm F435W} - m_{\rm F625W} \rangle_0).
$$
The resulting average value is $\langle m_{\rm
F625W,0}\rangle=15.4\pm0.3$.  This allows us to make an estimate of
the effective distance to the location of the majority of the
RCGs. The resulting distance modulus is $m-M=14.7\pm0.3$, which
translates into a distance $D=8.7\pm0.9$ kpc, in agreement with
recent distance estimates to the Galactic Center
\citep{eisenhauer05,ghez08}. The average extinction in the {\em HST}/ACS
field is $A_{F625W}\sim4$ and the extinction ranges from $\sim 3$ to $\sim5$.

\section{Constraints on the stellar density distribution in the \1p5} \label{sec_density}

The shape of the clump formed by the RCGs in Fig.~\ref{cmd_total}
suggests that their position is still affected by additional
extinction -- the distribution of the RCGs appears a bit elongated in the direction of the
reddening. This shows that the extinction varies even on angular scales
smaller than $0.96^\prime$. Note that this effect is almost absent
in regions with smaller total extinction -- after correction for
extinction,
the RCGs form a quite compact locus without an extension in the
reddening direction \cite[see e.g.][]{sumi04}.

However, apart from the possible elongation connected with any
residual reddening effects, the RCGs define a relatively narrow clump,
which can be used to constrain the distribution of stars
along the line of sight toward the \1p5.

\subsection{Model of the stellar distribution} \label{sec_model}

In any direction within the Galaxy stars are distributed over a range
of distances along the line of sight.  In the regions far away from the GC,
the dominant stellar component is the stellar disk, while at small
angular distances from the GC the majority of the stellar mass is
provided by the Galactic bulge and/or bar, nuclear stellar disk or nuclear
stellar cluster components. In our subsequent analysis we will
consider only three components of the stellar distribution: the
Galactic bulge, Galactic disk and nuclear stellar disk. We will adopt the distance to the Galactic Center of 8 kpc, which is some intermediate value between measurements of \cite{eisenhauer05} and \cite{ghez08}. The nuclear
stellar cluster is important only for the very central
parts of the Galaxy, at distances less than 10--20 arcmin \cite[see
e.g.][]{launhardt02}, and, therefore we will not consider it here. We use
the following simplified models of the three stellar components. Here $r$ is
the Galactocentric distance, $z$ is the height above the Galactic
plane.

For the stellar disk density we use (from \citealt{revnivtsev06})
$$
\rho_{\rm disk}=2.5~{\rm M_\odot ~pc^{-3}}  \times
$$
$$
\exp(-(3.0{\rm kpc}/r)^3-(r/2.5{\rm kpc})-|z/130{\rm pc}|).
$$

The normalization of this component was chosen to yield an
integrated stellar disk of $10^{10}~ M_\odot$, which is approximately
equivalent to the measured stellar density  in the solar
neighborhood ($0.045 M_\odot$
pc$^{-3}$,~e.g.  \citealt{robin03}).
As we will see below, details of the
stellar disk model are not important because of its small contribution
to the total mass budget in the \1p5.

For the Galactic bulge component we used the simple analytic form from
\cite{dehnen98} and \cite{grimm02}
$$
\rho_{\rm bulge}=1.09~{\rm M_\odot~pc^{-3}} \left( {\sqrt{r^2+(z/0.6)^2}}\over{1{\rm kpc}}\right)^{-1.8}\times
$$
$$
\exp\left(-{{r^2+(z/0.6)^2}\over{1.9 {\rm kpc}^2}}\right).
$$
The adopted total mass of the Galactic bulge is
$1.3\times10^{10}~M_\odot$ \citep{dwek95}.

We assumed that the nuclear stellar disk (NSD) has the following density distribution
$$
\rho_{\rm NSD}=\rho_d r^{-\alpha}e^{-|z|/0.045 {\rm kpc}},
$$
where $\rho_d=300 M_\odot$ pc$^{-3}$. At $r<0.12$ kpc, the slope
$\alpha=0.1$, at $0.12$ kpc $<r<0.22$ kpc, $\alpha=3.5$, and at $r>0.22$ kpc,
$\alpha=10$ \citep{launhardt02}.
The total mass of the NSD is thus $1.4\times10^9 M_\odot$. In reality
this quantity is uncertain by about 50\%
\citep{launhardt02}.

\begin{figure}[htb]
\includegraphics[width=\columnwidth]{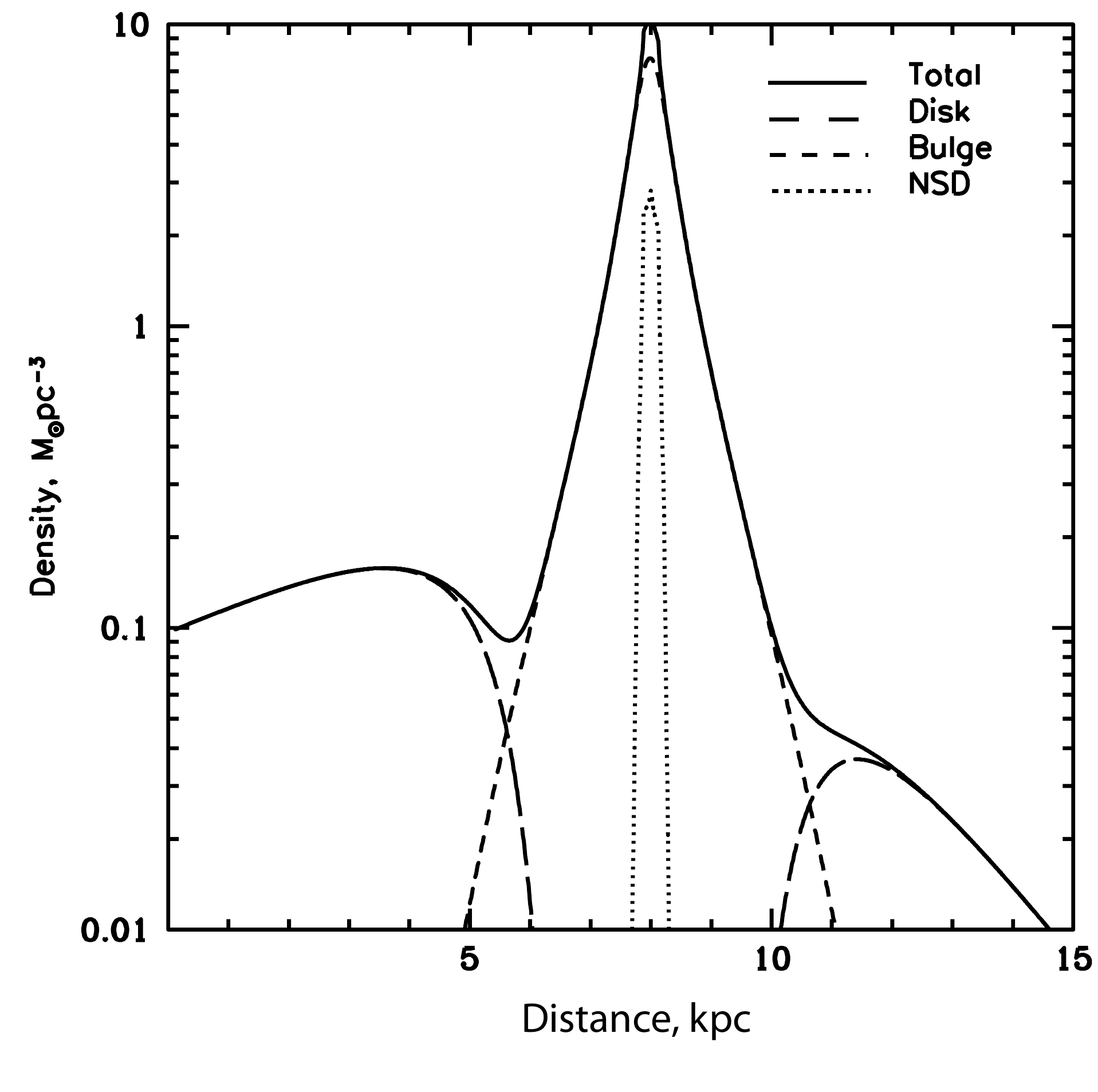}
\caption{Distribution of the density in the different adopted stellar
Galactic components along the line of sight in the direction of the
\1p5, i.e.~towards $(l^{II},b^{II})=(0.1^\circ,-1.42^\circ)$.}
\label{density_distribution}
\end{figure}

The resulting assumed distribution of the stellar density along the
line of sight in the direction of the \1p5 is presented in
Fig.~\ref{density_distribution}.  For the adopted model, the total
surface mass density in this direction is $4.6\times10^{4}~M_\odot$
sq.arcmin$^{-1}$, while the bulge component contributes
$3.9\times10^{4}~M_\odot$ sq.arcmin$^{-1}$, the stellar disk
contributes $2.3\times10^{3}~M_\odot$ sq.arcmin$^{-1}$, and the
nuclear stellar disk contributes $4.2\times10^{3}~M_\odot$
sq.arcmin$^{-1}$.

\subsection{Comparing the shape of the stellar model distribution with HST observations} \label{sec_distribution}

\begin{figure}[htb]
\includegraphics[width=\columnwidth,bb=14 128 581 750,clip]{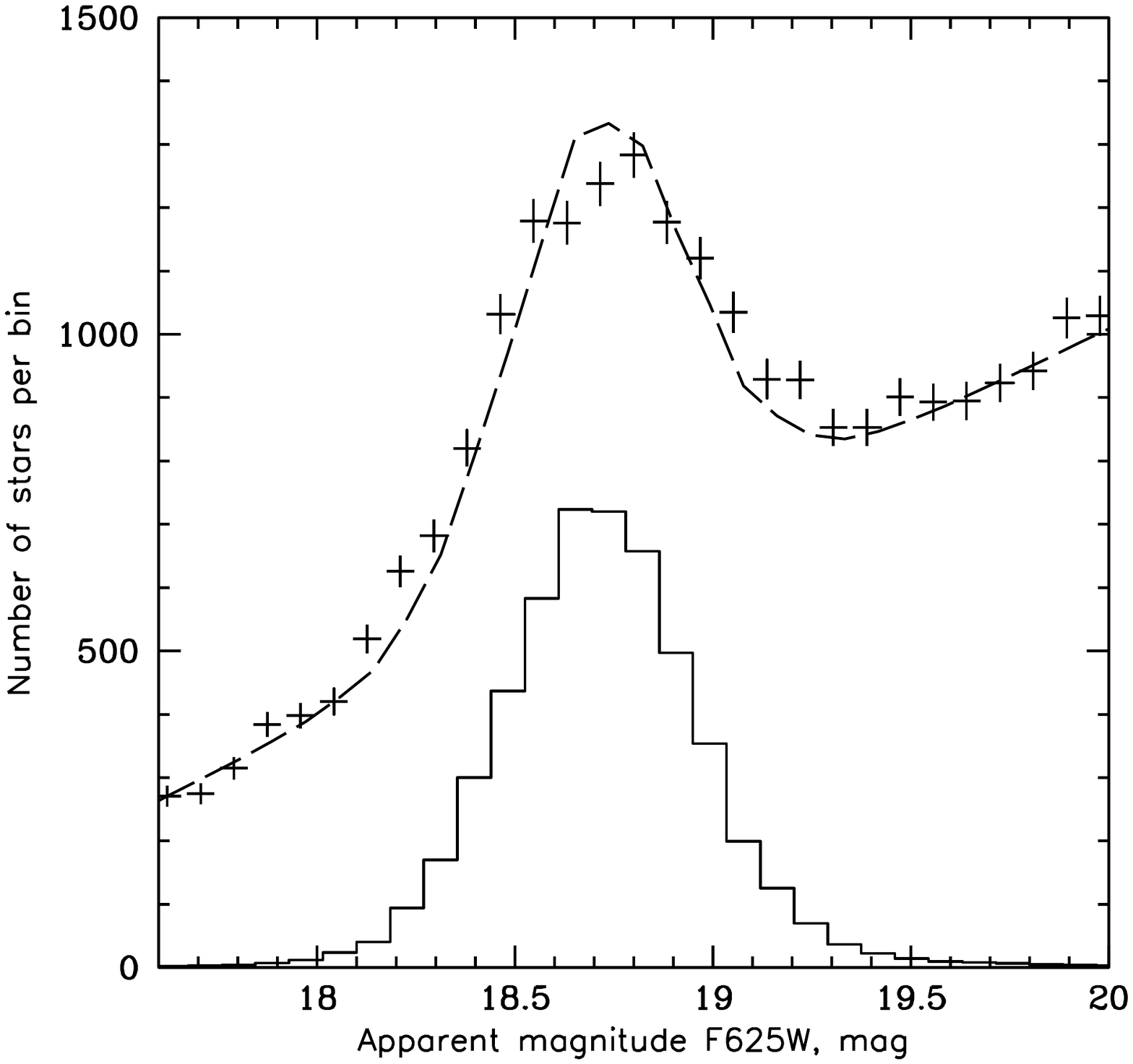}
\caption{Crosses - observed differential distribution of the number of stars
in the {\em HST}/ACS
field with color $m_{\rm F435W}-m_{\rm F625W}>2.5$ as a function of their apparent F625W magnitude
after correcting all magnitudes to a common value of the extinction
that corresponds to some arbitrary chosen position in the field of
view (see color-magnitude diagram on Fig.~\ref{cmd_total}). The size of
the magnitude bins is 0.085 mag. Solid histogram - modeled
differential distribution of RCGs as a function of their apparent
F625W magnitude in the field of our study. The dashed line shows a
part of the same distribution but with some linear function added to
match the surrounding magnitude bins.}
\label{mags_distribution}
\end{figure}

By choosing some fiducial number of the RCGs per unit mass of the stellar
population, we can use these stellar distribution models together with
the intrinsic magnitude distribution of RCGs (Fig.~\ref{distr_abs})
and the derived extinction in our field, to construct the differential
distribution of RCGs as a function of their apparent magnitude in the
F625W band and compare this prediction with the data. The model
distribution and the observations (distribution of photometered
stars from Fig.~\ref{cmd_total} with colors $m_{rm F435W}-m_{\rm
F625W} > 2.5$) are shown in Fig.~\ref{mags_distribution}. Here we
assumed that all (or the majority of) stars are located behind the
extinction region. This assumption is supported by studies of the
distribution of the interstellar extinction along the line of sight by
e.g. \cite{drimmel03} or \cite{marshall06}. See also Fig.~1 in
\cite{vandenberg09}, which shows that according to these models most
of the absorbing material is closer than $\sim$6 kpc, whereas
Fig.~\ref{density_distribution} in this paper shows that the bulk of
stars lies further away than $\sim$6 kpc.

Figure ~\ref{mags_distribution} shows the model
distribution of stars in the studied region (histogram and dashed line). It was obtained 
via convolution of the model of the stellar mass distribution over the line of 
sight with the distribution (see Fig.\ref{density_distribution}) of RCGs over 
their absolute magnitudes (see Fig.\ref{distr_abs}), assuming the denisty 
of the RCGs in their gaussian component  $N_{\rm RCG}/M_*\sim 1.6\times 10^{-3} ~M^{-1}_{\odot}$. 
This normalization value was chosen to fit the observed points. This RCG density 
number in general looks reasonable (see e.g. \citealt{lopez02}), however, a more 
quantitative comparison of this number with those obtained in different observational
 and theoretical works requires more accurate treating of different selection effects. 
The figure shows that the 
model distribution very closely follows the
measured points (crosses) around the corresponding peak of the
color-magnitude diagram. The linear function that is added to make the
model distribution match with the magnitude bins that surround the
observed peak could originate from red giants and asymptotic giant
branch stars that also lie in this part of the color-magnitude diagram,
but do not belong to the red clump (see also \citealt{alves00}). 

There are some other subtle
differences to note. In particular, the model (which takes into
account the intrinsic dispersion of the RCGs brightnesses at the level
of $0.19$ mag) has a Gaussian width of the main peak of $0.23$
mag. However, the Gaussian width of the RCG
peak is $0.28\pm0.01$ in the \1p5. This is somewhat higher than we
might expect from our model. 

We should keep in mind though that
there may be several complications that could result in the observed
additional broadening. For example, it is likely that we have not
completely corrected for all the extinction in our field due to its
variations on small angular scales (see the discussion of the shape of the
red clump in Fig.~\ref{cmd_total}); with the adopted technique based
on RCGs, these variations are difficult to account for, because the
density of RCGs is too low to use smaller spatial bins. Residual
variations of the total extinction in our field with an amplitude
$\Delta A_{\rm F625W}\sim 0.16$ mag can result in the observed
broadening of the RCG distribution. 

It is also possible that the
intrinsic (i.e. approximately extinction-corrected) brightness of the RCGs in the
Galactic bulge has a slightly higher scatter than that in the Solar
vicinity, which we used to compute the model
distribution. 

It is also possible that the extra broadening is the
result of our oversimplified picture that all the extinction is
located in front of the majority of the Galactic bulge RCGs. If some
residual extinction, distributed {\em within} the Galactic bulge, is
present in our field, it will similarly result in a slight widening of
the RCG peak in the differential number-counts plot.

Finally it could
be that the adopted model for the stellar density distribution of the
bulge oversimplifies the true distribution, resulting in a density
peak that is too narrow. However, our tests to vary the parameters of
the stellar model showed that the latter is unlikely.

Summarizing the above findings, we conclude that the shape of the
adopted model of the stellar density distribution in the Galaxy
adequately describes the observed distribution in the \1p5.

\subsection{Stellar surface brightness in the near-infrared} \label{sec_spitzer}

\begin{figure}[htb]
\includegraphics[width=\columnwidth]{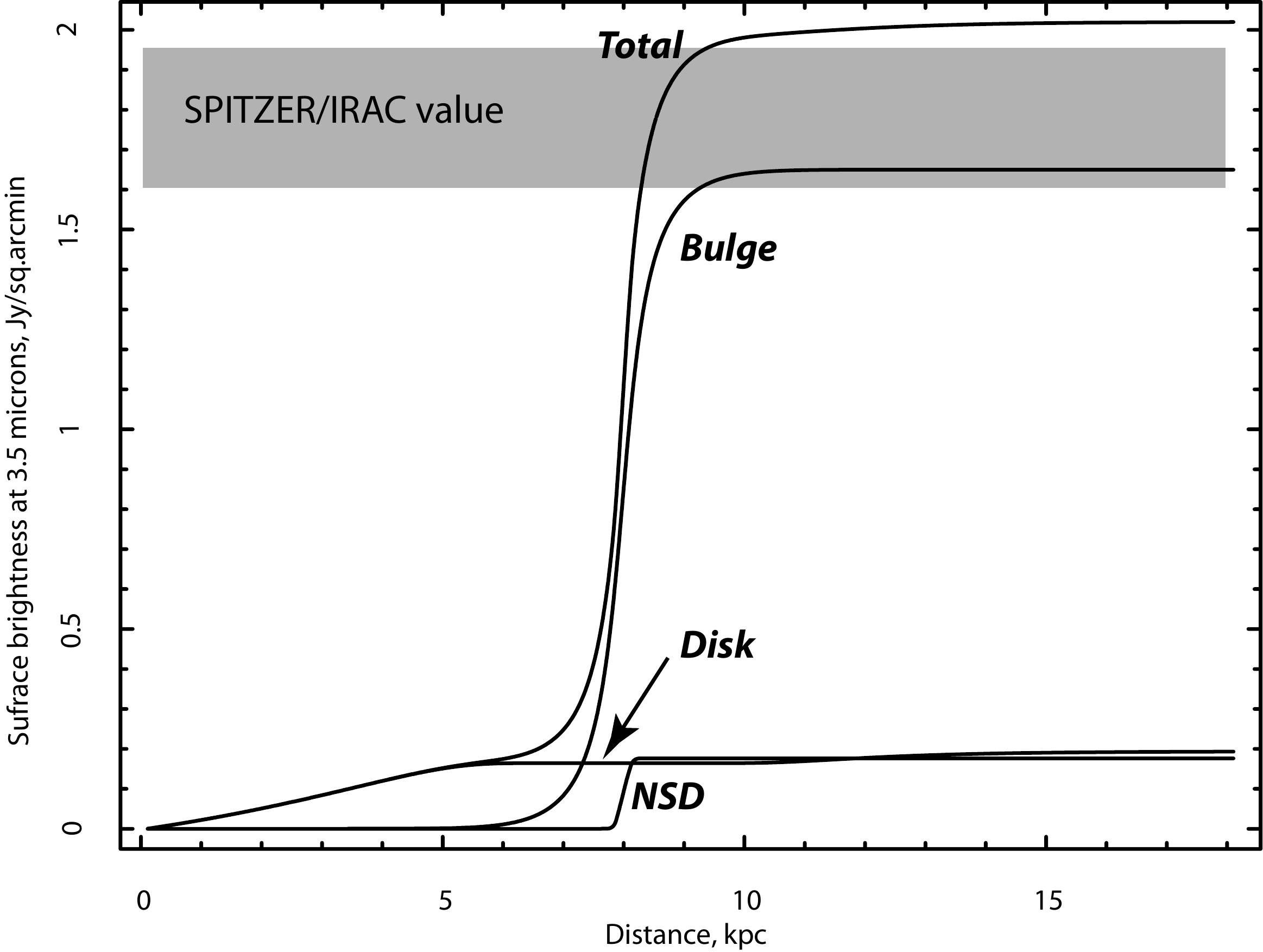}
\caption{Cumulative stellar surface brightness at 3.5 $\mu$m in the direction of
the \1p5 as predicted by the adopted stellar mass model, decomposed in
the three components (``Bulge'', ``Disk'' and ``NSD'') as well as
their total. The measured surface brightness as measured
by {\em Spitzer}/IRAC at 3.6 $\mu$m is shown by the gray area.}
\label{cumbrightness}
\end{figure}

Another important check of the adopted stellar mass distribution in
the direction of the \1p5 is the total surface brightness provided by
stars. We took the value of the total luminosity density of the
Galactic bulge as measured by \cite{dwek95} at a wavelength 3.5
$\mu$m -- $4.19\times10^{28}$ erg s$^{-1}$ Hz$^{-1}$ -- and assigned
it to our adopted total mass of stars in the Galactic bulge --
$1.3\times10^{10}~M_\odot$ -- to calculate a specific infrared luminosity
density per unit mass. Then we integrated the surface brightness
provided by all stars in our model (assuming the same luminosity-to-mass ratio for all stellar components). We point out that due to
differences in age and/or composition compared to the bulge, the true
specific luminosity density at 3.5 $\mu$m could be slightly different
for the (nuclear) disk, although their contributions
to the total surface brightness are relatively small (see below).  The
result of this integration is shown in Fig.~\ref{cumbrightness}. The
observed infrared surface brightness at 3.6 $\mu$m in our field
is $21\pm2$ MJy/sr, or $1.78\pm0.18$ Jy sq.arcmin$^{-1}$, as measured by
{\em Spitzer}/IRAC
\cite[see][]{revnivtsev09}. It is seen that the adopted model
correctly predicts
the observed surface brightness. The majority of the 3.5
$\mu$m flux is provided by the stellar population in the Galactic
bulge ($\sim$82\%) with small contributions from the stellar disk
($\sim$9\%) and the nuclear stellar disk ($\sim$9\%).  The dominance
of the Galactic bulge stars to the total stellar mass budget in the field
is also evident from Fig.~\ref{image}. The vast majority of stars in
the high resolution {\em HST}/ACS image are faint and have a
yellow/red color, which shows that they are of a relatively low mass and
are located at a large distance from us. This is exactly what is expected
for a stellar population of the Galactic bulge with an age of $\sim$10
Gyr.

\section{Conclusions} \label{sec_conclusion}

The study of populations of Galactic X-ray sources requires knowledge
of the properties of the underlying population of normal stars,
because only then one can estimate the statistical properties of X-ray
sources and their systematic dependencies on the age, metallicity,
density etc. of the stellar population. One of the largest projects
performed recently by the {\em Chandra X-ray Observatory} is the
ultra-deep observation of a region close to the Galactic Center - the
\1p5. This observation provides us
with a wealth of data on the X-ray population in this area. To connect
these properties to the properties of normal stars we must study
this area in the optical and near-infrared bands.

We analyzed {\em HST}/ACS images of the central 6.6\arcmin
$\times$ 6.6\arcmin\, field of the \1p5 in the F435W and F625W bands and
found the following:

\begin{itemize}
\item Interstellar extinction in the direction of the \1p5 is $\langle
A_{\rm F625W}\rangle \sim 4$ and variable on an angular scale of
$\sim$1\arcmin. Variations of the extinction in this area on larger
scales were previously noticed by \cite{dutra03} ($4^\prime$ scales),
\cite{revnivtsev09b} ($1.8^\prime$ scales),
\cite{vandenberg09} ($1.6^\prime\times3.2^\prime$ scales).

\item The color-magnitude diagram corrected for extinction, as determined in
angular bins of $0.96^\prime\times0.96^\prime$, suggests that there is
still residual extinction on smaller scales at the level of $A_{\rm
F625W}\sim 0.2$ mag or less. This would be difficult to measure with RCGs
due to their decreasing numbers.

\item We determined the extinction law (in particular, the ratio of the 
total-to-selective extinction $\Delta \langle m_{\rm F625W} \rangle
/\Delta \langle m_{\rm F435W}-m_{\rm F625W}\rangle$) in the direction
of the \1p5  -- $1.3\pm0.1$ -- and found that it is significantly different from the
canonical one $\sim1.9$.
The discrepancy shows a similar trend as that determined by
various authors using OGLE data of the bulge
\cite[e.g.][]{popowski00,udalski03,sumi04}. This suggests that we
might expect deviations from the standard correspondence between the
photoabsorption column $n_HL$, measurable in the X-ray band, and
extinction in the optical/NIR bands. This likely differing X-ray (vs. IR)
absorption in the bulge region will affect derived intrinsic X-ray spectra
and thus (in some cases) the inferred source types (e.g., with less X-ray
 absorption for a given color excess, soft spectral components will be
 decreased, etc.).

\item We compared a model of the stellar distribution along the line of
sight towards the \1p5 with the observed distribution of red clump
giants in the color-magnitude diagram. We show that the
adopted model adequately describes the data. In the model, the
majority of the stellar mass in the direction of the \1p5 is provided
by Galactic bulge stars with only a small contribution from the stellar disk
and nuclear stellar disk components ($\sim5$\% and $\sim$9\%,
respectively). Small discrepancies between the model and the data
could be the result of variations in the extinction on small angular
scales ($\lesssim$1\arcmin), differences between the local and bulge
luminosity functions of red clump giants, uncertainties in the
distribution of the extinction along the line of sight, or an
oversimplified model of the stellar mass profile of the bulge.

\item
We showed that the adopted stellar model predicts an integrated
3.5 $\mu$m surface brightness of this part of the Galaxy that is only
13\% higher than the surface brightness actually measured by {\em
Spitzer}/IRAC in the 3.6 $\mu$m band.
\end{itemize}

\begin{acknowledgements}
The authors thank Annamaria Donnarumma and Maxim Markevitch for their
help in obtaining Magellan observations of the field.  This research
made use of data obtained from the High Energy Astrophysics Science
Archive Research Center Online Service, provided by the NASA/Goddard
Space Flight Center. This work was supported by grants of the Russian
Foundation of Basic Research (07-02-01051, 07-02-00961-a, 08-08-13734, 07-02-01004, 08-02-00974, NSh-5579.2008.2) and programs of Presidium of RAS P04 and OFN-17. MvdB was supported in part by
STScI/HST grant HST-GO-10353.01. Results in this paper are based on
observations made with the NASA/ESA Hubble Space Telescope, obtained
at the Space Telescope Institute which is operated by the Association
of Universities for Research in Astronomy, Inc. under the NASA
contract NAS 5-26555.
\end{acknowledgements}

\end{document}